\documentclass{optica-article}

\journal{opticajournal} 

\articletype{Research Article}

\usepackage{lineno}

\usepackage{comment}

\newcommand{\ngr}{n_{\rm g}}
\newcommand{\LL}{L_{\rm L}}
\newcommand{\LS}{L_{\rm S}}

\begin{document}

\title{Efficient light couplers to topological slow light waveguides in valley photonic crystals}

\author{Hironobu Yoshimi,\authormark{1,2,*} Takuto Yamaguchi,\authormark{1,2} Satomi Ishida,\authormark{1,2} Yasutomo Ota,\authormark{3} and Satoshi Iwamoto\authormark{1,2,**}}

\address{\authormark{1}Research Center for Advanced Science and Technology, The University of Tokyo, 4-6-1 Komaba, Meguro-ku, Tokyo 153-8904, Japan\\
\authormark{2}Institute of Industrial Science, The University of Tokyo, 4-6-1 Komaba, Meguro-ku, Tokyo 153-8505, Japan\\
\authormark{3}Department of Applied Physics and Physico-Informatics, Faculty of Science and Technology, Keio University, 3-14-1 Hiyoshi, Kohoku-ku, Yokohama-shi, Kanagawa 223-8522, Japan
}

\email{\authormark{*}hyoshimi@iis.u-tokyo.ac.jp} 
\email{\authormark{**}iwamoto@iis.u-tokyo.ac.jp}


\begin{abstract*}
We numerically and experimentally demonstrate efficient light couplers between topological slow light waveguides in valley photonic crystals (VPhCs) and wire waveguides. By numerical simulations, we obtained a high coupling efficiency of $-0.84$ dB/coupler on average in the slow light regime of a group index $\ngr = 10 - 30$. Experimentally, we fabricated the couplers in a Si slab and measured the transmitted power of the devices. We realized a high coupling efficiency of approximately $-1.2$ dB/coupler in the slow light region of $\ngr = 10 - 30$, which is close to the result from the numerical simulations. These demonstrations will lay the groundwork for low-loss photonic integrated circuits using topological slow light waveguides.
\end{abstract*}

\section{Introduction}

Optical waveguides in topological photonic crystals have gathered considerable attention because of their potential for robust light propagation \cite{ozawa2019topological, iwamoto2021recent}. Especially, valley photonic crystal (VPhC) waveguides enable efficient light propagation through sharp waveguide bends. The waveguides can be implemented in normal dielectrics such as Si, and thus are compatible with the photonic integrated circuit (PIC) technologies \cite{shalaev2019robust, he2019silicon, ma2019topological, yamaguchi2019gaas, arora2021direct, arregui2021quantifying}. 
These features can be applied to the further dense integration and functionalization of the PICs.
In recent years, novel photonic devices and phenomena based on topological VPhC waveguides have been demonstrated, such as compact optical switches \cite{wang2022ultracompact}, ring lasers \cite{noh2020experimental, gong2020topological}, quantum photonic circuits \cite{chen2021topologically}, chiral light-matter interactions \cite{mehrabad2020chiral, mehrabad2023chiral}, and second-harmonic generation \cite{lan2021second}.

Recently, topological slow light waveguides using a bearded interface formed in VPhCs have been proposed and demonstrated \cite{yoshimi2020slow, yoshimi2021experimental}, providing an effective way to suppress large bending losses in the conventional PhC slow light waveguides \cite{assefa2006transmission}. This slow light edge state can also be used for enhancing light-matter interactions with topological protection \cite{kuruma2022topologically, miyazaki2021lasing, xie2021topological}. When introduced into the various novel VPhC devices \cite{mehrabad2020chiral, mehrabad2023chiral, chen2021topologically, wang2022ultracompact, noh2020experimental, gong2020topological, lan2021second}, the topological slow light waveguides in VPhCs could lead to further advanced photonic technologies.

For implementing the VPhC slow light waveguides in PICs, efficient light coupling between the slow light waveguide in VPhCs and a wire waveguide is essential. Previously, for waveguides in photonic topological insulators \cite{barik2018topological}, efficient light coupling from a wire waveguide was demonstrated numerically and experimentally \cite{kagami2020topological}. For the VPhC waveguides formed at the zigzag interface, a few ideas on coupling structures for fast light modes have been investigated \cite{shalaev2019robust, chen2022efficient}. 
In a coupler structure with a line defect, over $95\%$ coupling efficiency was demonstrated using numerical simulations of two-dimensional (2D) models \cite{chen2022efficient}.
For the edge states formed at the bearded interface in VPhCs, a coupler structure with a line defect has been examined \cite{mehrabad2020chiral, chen2022efficient}. Although a high coupling efficiency of $\sim 94\%$ was calculated in 2D simulations \cite{chen2022efficient}, the average transmittance in the topological band calculated by the three-dimensional (3D) simulations is only $\sim 31\%$ \cite{mehrabad2020chiral}. Furthermore, the coupling efficiency was not measured experimentally.
One of the reasons for the low coupling efficiency into the bearded interface will be the existence of the glide plane symmetry across the interface \cite{mock2010space}. The glide plane symmetry leads to the complex field distributions of the modes, making it challenging to realize efficient light coupling from a simple wire waveguide \cite{mahmoodian2017engineering, sollner2015deterministic}. Design of an efficient light coupler in the slow light regime based on the 3D simulations and its experimental demonstrations are desired for the practical implementation of the VPhC slow light waveguides.

Here, we numerically and experimentally demonstrate an efficient light coupler between a slow light waveguide in VPhCs and a wire waveguide. We investigated a coupling structure which is formed by filling air holes adjacent to the interface between topologically distinct VPhCs. A coupling efficiency of $-0.84$ dB/coupler was obtained on average in the slow light region of $\ngr = 10 - 30$ by the 3D simulations. We fabricated Si-based couplers and experimentally observed a high coupling efficiency of approximately $-1.2$ dB/coupler in the slow light regime of $\ngr = 10 - 30$, which is close to the numerical result. These results provide an important step for realizing PICs with topologically protected slow light waveguides.
We note that the coupling efficiency of a similar coupler structure for a topological slow light waveguide is reported in the supplemental material of a recently published study \cite{rosiek2022observation}. The study reports the transmittance of approximately $-0.6$ to $-1.4$ dB/coupler in the slow light regime of $\ngr = 10 - 30$ in experiment, which is close to our result.
On the other hand, the report does not investigate the dependence of the efficiency on the coupler length or in-gap modes in the coupling region, which are discussed in this paper.

\begin{figure*}[tb]
\centering\includegraphics[width=\linewidth]{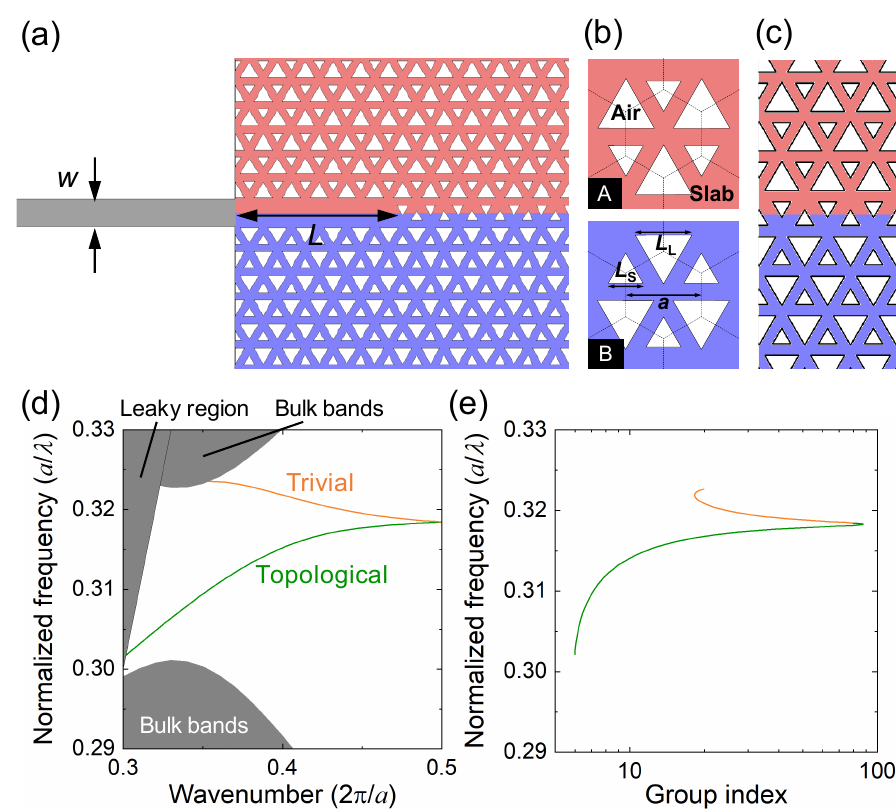}
\caption{
(a) Schematic of the investigated coupler. 
(b) Schematic of the unit cells of VPhCs A and B.
(c) Top view of the bearded interface between two topologically distinct VPhCs.
(d) Dispersion curves of the in-gap states formed at the bearded interface in Fig. 1 (c). The green and orange curves represent topological and trivial states, respectively. The simulations are performed by the 3D PWE method.
(e) Group index spectra for the in-gap states shown in (d).
}
\end{figure*}

\section{Device design and simulations}

Figure 1(a) shows a schematic of the investigated coupler. The wire waveguide is shown in gray and topologically distinct VPhCs in a dielectric slab are shown in red and blue. Air holes adjacent to the interface between the two VPhCs are filled with the dielectric. This region is indicated by the arrows in Fig. 1(a), and its length is denoted by $L$.
This coupler structure is similar to the one discussed in the reference \cite{mehrabad2020chiral}; however, the structure at the termination of VPhC is different. While in the structure \cite{mehrabad2020chiral} the VPhCs are terminated so that no air hole is placed on the end face, the end face in the investigated structure is placed by crossing the center of holes \cite{chen2022efficient, rosiek2022observation}.
Unit cells of the VPhCs are shown in Fig. 1(b). Equilateral triangular air holes are patterned in a honeycomb lattice with a period of $a = 500$ nm and a slab thickness of $d = 220$ nm. The refractive index of the slab is set to 3.48 in the following simulations. The side lengths of the triangles are denoted by $\LL$ and $\LS$. For $\LL = \LS$, the PhCs possess $C_{6v}$ point group symmetry and the photonic bands support gapless Dirac cones at K and K' points in the momentum space. On the other hand, when $\LL \neq \LS$, the degeneracy at the Dirac points is lifted and the topological bandgap opens. Although the unit cells A and B possess the same energy band structures, their band topologies are distinct \cite{shalaev2019robust}. In the following, $\LL = 1.2a/\sqrt{3}$ and $\LS = 0.75a/\sqrt{3}$ are used. Figure 1(c) shows a bearded interface between the topologically distinct VPhCs. A topological edge state appears around the interface originating from the distinct band topologies in VPhCs A and B. The dispersion curves of the in-gap states formed at the bearded interface are shown in Fig. 1(d), calculated by the 3D plane wave expansion (PWE) method. The bearded interface supports two curves and they are degenerated at the Brillouin zone (BZ) edge due to the glide plane symmetry of the interface \cite{yoshimi2020slow, mock2010space}. The band assignments of the two curves are discussed in our previous papers \cite{yoshimi2020slow, yoshimi2021experimental}, and the higher (orange) and the lower (green) frequency bands correspond to the trivial and topological states, respectively. Thus, the modes in the lower frequency curve benefit from the topological protection and exhibit efficient light propagation through sharp wavaguide bends. A prominent feature of this dispersion curve is the presence of a slow light region within the topological bandgap below the light line. The group index spectra for the in-gap states are shown in Fig. 1(e). The orange and green curves respectively represent the trivial and topological states. We can confirm the slow light region near the BZ edge and the group index approaches 100. We note that the maximum group index of the modes in the topological band can be increased to over 100 by employing optimum air hole sizes such as $\LL = 1.3a/\sqrt{3}$ and $\LS = 0.6a/\sqrt{3}$.

\begin{figure}[tb]
\centering\includegraphics[width=0.8\linewidth]{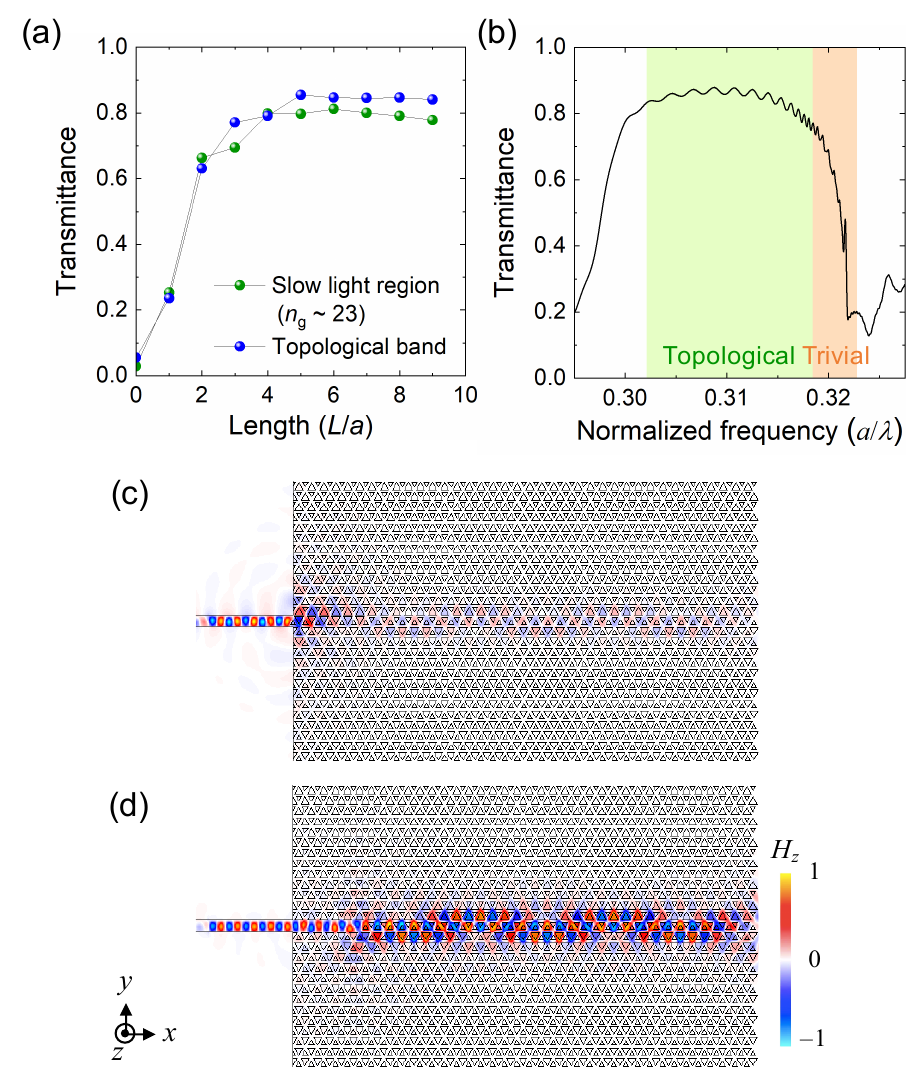}
\caption{
(a) Calculated transmittance of the investigated coupler as a function of the length $L$. Green dots correspond to the slow light regime ($\ngr \sim 23$) in the topological band. Blue dots represent the average in the whole topological band. (b) Transmission spectrum of the investigated coupler with $L = 6a$. The green and orange regions indicate the frequency bands of the topological and trivial states, respectively. (c) and (d) Distributions of the magnetic field $H_z$ when the light propagates through (c) the butt coupling structure ($L = 0$) and (d) the investigated coupler ($L = 6a$). The group index of the light in the VPhC slow light waveguide is $\sim 23$. The simulations are performed by the 3D FDTD method.
}
\end{figure}

Subsequently, we calculated light transmittance through the investigated coupler shown in Fig. 1(a). Figure 2(a) shows the simulated transmittance as a function of the length $L$ by the 3D finite difference time domain (FDTD) method. The width of the wire waveguide is set to $w = 480$ nm. The green dots represent the transmittance of the slow light regime in the topological band with the moderate group index of $\ngr \sim 23$. The blue dots indicate the average transmittance over the entire frequency band of the topological edge state, which is represented by the green curve in Fig. 1(d). For the length $L = 0$ (butt coupling structure), the transmittance is only less than 0.1 ($-10$ dB). In the slow light region (green dots), when the length $L$ becomes longer, the transmittance increases and becomes larger than 0.8 ($-0.97$ dB) for $L = 6a$. For the length $L > 6a$, the transmittance gradually decreases. This is because the waveguide mode in the coupler region lies above the light line and leaks in the out-of-plane directions. Therefore, the transmittance exhibits a maximum value at an optimum length. Dispersion curves of the in-gap modes at the coupler region are discussed later in Fig. 3.
Distribution of the magnetic field $H_z$ of light propagating through the butt coupling structure is shown in Fig. 2(c). The group index of the mode in the VPhC slow light waveguide is $\sim 23$. Most of the input light is scattered at the butt coupling structure and the intensity in the VPhC slow light waveguide is quite low. Distribution of the magnetic field $H_z$ of light propagating through the investigated coupler for $L = 6a$ is shown in Fig. 2(d). Although the mode distributions of the topological slow light waveguide and the wire waveguide are significantly different, they can be efficiently coupled when the region with filled air holes is inserted. We note that the twists in the field distribution of the slow light waveguide are simply because the real part of the amplitude is shown here. When we plot the intensity of light, these twists are not observed.
Figure 2(b) shows the transmission spectrum of the coupler when the length $L$ is $6a$. The green and orange shaded regions represent the frequency bands of the topological and trivial states, respectively. For most of the frequency band of the topological edge state, the transmittance is larger than 0.8 ($-0.97$ dB). Especially, in the slow light regime of $\ngr = 10 - 30$, the average transmittance is $-0.84$ dB/coupler, exhibiting efficient light coupling from the wire waveguide to the topological slow light waveguide in VPhCs.


Next, we discuss the dispersion curves of the in-gap modes in the coupler region. Figure 3(a) shows a schematic of the coupler region. Topologically distinct VPhCs are shown in red and blue. Air holes adjacent to the interface between the two VPhCs are filled with Si, as we explained in Fig. 1(a). Dispersion curves of the in-gap states formed at the coupler region are shown in Fig. 3(b). Bulk modes and a light cone are shown by gray and light blue shaded regions, respectively. In the investigated coupler, the modes within the topological bandgap are utilized. Focusing on the dispersion curve within the topological bandgap, the modes are located in the light cone. Thus, the power of these modes leaks into the air-cladding region from the slab as propagation. 
When the length of the coupler region becomes longer, this power leakage is prominent and the transmittance of the investigated coupler is decreased.
The simulation results show that the optimum coupler length in the slow light regime is $\sim 6a$, as we have shown in Fig. 2(a).

\begin{figure}[tb]
\centering\includegraphics[width=0.8\linewidth]{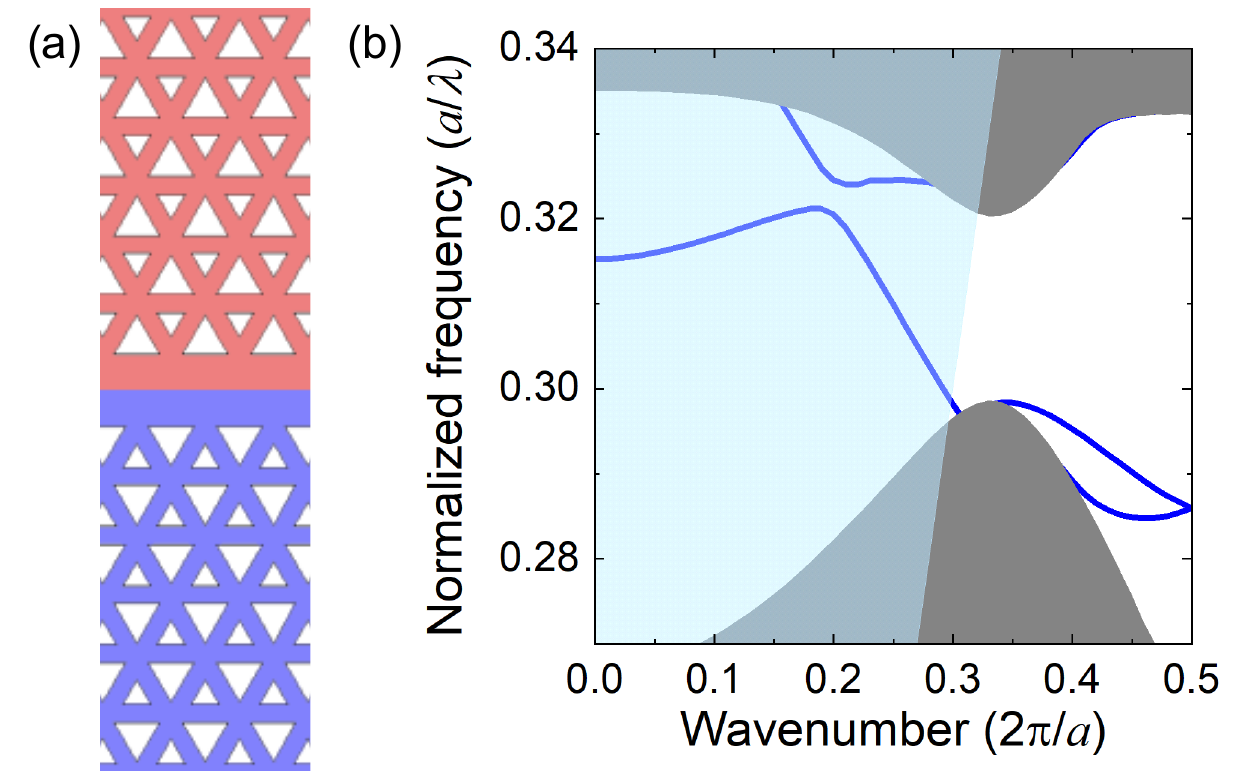}
\caption{
(a) A schematic of the coupler region. (b) Dispersion curves of the in-gap modes in the coupler region (blue curves). Gray and light blue regions represent bulk modes and a light cone, respectively.
}
\end{figure}

\begin{figure}[tb]
\centering\includegraphics[width=\linewidth]{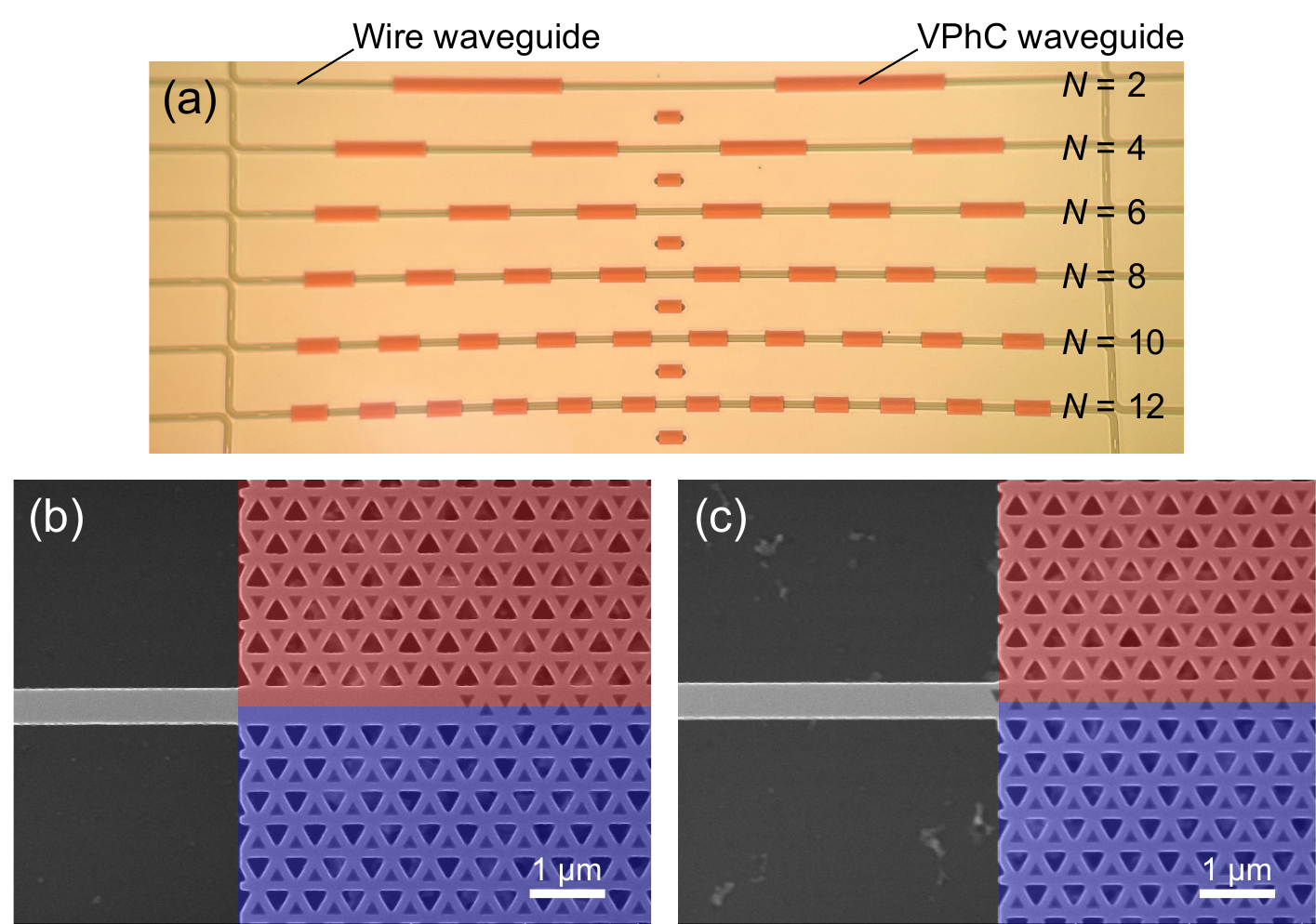}
\caption{
(a) An optical microscope image of the fabricated devices with $N$-section VPhC waveguides ($N$ = 2, 4, 6, 8, 10, and 12). SEM images of (b) the investigated coupler and (c) the butt coupling structure. 
}
\end{figure}

\section{Device fabrication and optical characterization}

We fabricated the investigated couplers in an air-bridge 220 nm-thick Si slab with a combination of electron beam lithography, reactive ion etching, and vapor hydrogen fluoride etching.
An optical microscope image of the fabricated devices is shown in Fig. 4(a). We fabricated $N$-section VPhC waveguides ($N$ = 2, 4, 6, 8, 10, and 12). Each device has a total VPhC waveguide of 300 $\upmu$m and $2N$ couplers. These devices have the constant total waveguide lengths and different numbers of the couplers, and thus we can experimentally obtain the efficiency of the couplers by measuring the transmitted power of the devices. Figure 4(b) shows the scanning electron micrograph (SEM) image of the fabricated coupler. For comparison, we also fabricated butt coupling structures between VPhC slow light waveguides and the wire waveguides, as shown in Fig. 4(c). In addition, we fabricated a reference wire waveguide, which does not include VPhC waveguides. We note that the wire waveguide sections are air-bridged by Si wire supporters with low loss elliptical intersections \cite{fukazawa2004low}.

We characterized the coupling efficiency of the investigated structure by measuring the transmitted power of the devices. We input laser light into one facet of the device and measured the transmitted power from the other facet by an optical spectrum analyzer (OSA). We used a wavelength-tunable laser diode as a light source and obtained transmitted power spectra, as shown in Fig. 5(a). The spectra are normalized by the transmitted power of the wire waveguide which does not include VPhC waveguides. The green and orange regions respectively correspond to the frequency bands of the topological and trivial states. The assignment of the topological band and the slow light regime was performed by the group index measurement of the waveguide modes by using another device located close to the target devices, as reported in our previous paper \cite{yoshimi2021experimental}. We estimated that the slow light regime of $\ngr = 10 - 30$ in the topological band corresponds to the wavelengths $\lambda = 1537 - 1554$ nm. Figure 5(b) shows the measured transmittance in the slow light regime for the investigated coupler (blue) and the butt coupling structure (red), as a function of the number of couplers. Each point is averaged over the slow light regime of $\ngr = 10 - 30$. We can confirm that the transmitted power decreased as the nubmer of the couplers increased. Blue and red lines show linear regression lines for the transmittance of the investigated coupler and the butt coupling structure. From the slope of the lines, we calculated the coupling efficiencies of the two coupling structures. The coupling efficiencies for the investigated and butt couplers are $-1.21$ and $-2.64$ dB/coupler, respectively. The experimental coupling efficiency for the investigated coupler is close to the one from the numerical simulations. We also plotted a point as a star mark for the number of couplers 0 based on the recently reported propagation loss in the VPhC slow light waveguides with the group index of $\ngr \sim 20$ \cite{rosiek2022observation}.
The propagation losses obtained from the fitting lines for the coupling structures (y-intercepts of the lines) are close to the reported propagation loss.

\begin{figure}[tb]
\centering\includegraphics[width=\linewidth]{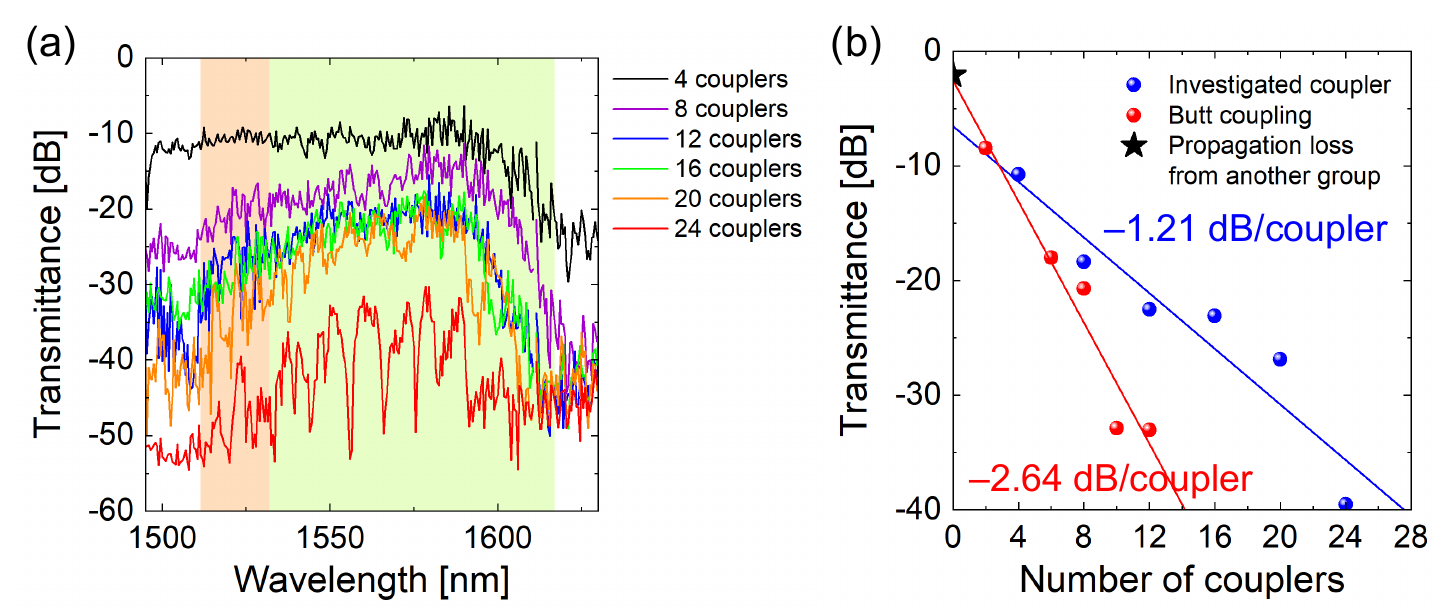}
\caption{
(a) Measured transmission spectra for the devices including 4 (black curve), 8 (purple), 12 (blue), 16 (green), 20 (orange), and 24 (red) couplers. The green and orange shaded regions represent the frequency band of the topological and trivial in-gap states, respectively. (b) Coupling efficiency of the investigated coupler (blue dots) and the butt coupling structure (red dots). Blue and red lines represent linear regression lines for the blue and red dots, respectively. A star mark indicates the propagation loss reported from another group \cite{rosiek2022observation}.
}
\end{figure}

\section{Conclusion}

In conclusion, we realized efficient light couplers to the topological slow light waveguide in VPhCs from a wire waveguide. Using the 3D numerical calculations, a high coupling efficiency of $-0.84$ dB/coupler is demonstrated in the slow light regime of $\ngr = 10 - 30$. In experiment, we fabricated Si-slab air-bridge couplers for the topological slow light waveguides and demonstrated efficient light coupling of approximately $-1.2$ dB/coupler in the slow light regime of $\ngr = 10 - 30$, which is close to the numerical results. These demonstrations are an important step toward PICs utilizing topological photonic waveguides \cite{iwamoto2021recent}, active topological photonic devices such as lasers and optical switches \cite{ota2019active, noh2020experimental, wang2022ultracompact}, topological quantum photonic devices \cite{chen2021topologically}, and enhancement of light-matter interactions and nonlinear optical phenomena in topological photonic crystals \cite{smirnova2020nonlinear, mehrabad2020chiral, mehrabad2023chiral, lan2021second}.


\section*{Funding}

This work is supported by JSPS KAKENHI (17H06138, 21J12323, and 22H00298), JST CREST (JPMJCR19T1), New Energy and Industrial Technology Development Organization (NEDO) and The Asahi Glass Foundation Research Grant Program.

\section*{Acknowledgments}
The authors thank Dr. F. Tian and Mr. N. Harada for fruitful discussions.

\section*{Disclosures}

The authors declare no conflicts of interest.

\bibliography{sample}






\end{document}